\documentclass[twoside,a4paper,11pt]{torus2012}
\usepackage{graphicx}
\usepackage{hyperref}
\usepackage{movie15}
\begin{document}
\pagenumbering{arabic}
\pagestyle{myheadings}
\thispagestyle{empty}
{\flushleft\includegraphics[width=\textwidth,bb=58 650 590 680]{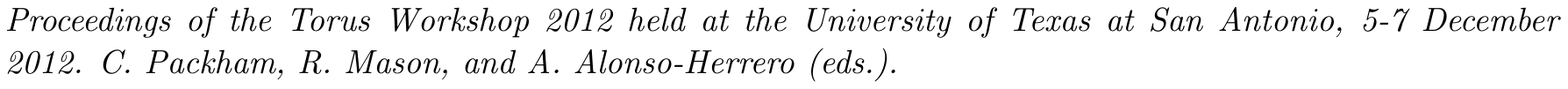}}
\vspace*{0.2cm}
\begin{flushleft}
{\bf {\LARGE
Mid-infrared Imaging of Nearby Radio-Loud AGN
%
}\\
\vspace*{1cm}
%
Eric S. Perlman$^{1}$, Matthew Merlo$^{1}$, Chris Packham$^{2}$, Rachel Mason$^{3}$, and 
Los Piratas.
%
}\\
\vspace*{0.5cm}
%
$^{1}$
Department of Physics \& Space Sciences, Florida Institute of Technology, 150 W. University Blvd., Melbourne, FL  32901, USA \\
$^{2}$
Department of Physics \& Astronomy, University of Texas at San Antonio, 1 UTSA Circle, San Antonio, TX  78249, USA\\
$^{3}$
Gemini Observatories, 670 N. A'ohoku Place, Hilo, HI 96720, USA
%
\end{flushleft}
%
\markboth{
Mid-IR Imaging of Radio-Loud AGN
}{ 
%
Perlman et al.
%
}
\thispagestyle{empty}
\vspace*{0.4cm}
\begin{minipage}[l]{0.09\textwidth}
\ 
\end{minipage}
\begin{minipage}[r]{0.9\textwidth}
\vspace{1cm}
\section*{Abstract}{\small

We present high spatial resolution MIR observations for several nearby radio loud active galactic 
nuclei (RLAGN), which were obtained using the Gemini North and South telescopes.  Of the six 
observed objects, we detected five in the Si-2 (8.7 microns) and Si-6 (12.3 microns) filters, of which 
two objects show some evidence of low level extended emission surrounding the unresolved nucleus.   
In Pictor A, we also obtained an image in Qs (18.3 microns) that has a flux of only half that seen in the 
Spitzer image, suggesting structure on arcsecond scales.  We also used the Si-6 (12.3 microns) flux 
measurement to investigate correlation between our MIR flux and xray luminosity and compare this to 
results  for AGN in general.  
This work also forms a basis for future high resolution imaging and spectroscopy of these objects. 
%
\normalsize}
\end{minipage}
\section{Introduction \label{intro}}

Active galactic nuclei (AGN) include many different object classes with drastically different properties, 
from ultra-luminous quasars to Seyfert galaxies to galaxies with powerful radio jets.  These diverse 
classes span as many as eight decades in luminosity, include objects with and without broad optical 
emission lines, and have varied SED shapes.  One of the major developments in the study 
of AGN was the proposal of a unified theory, which allows these vastly different observed properties to 
be explained by one structure, instead of a different type of system for each type of object.  The unified 
theory proposes a system with a super-massive black hole, accretion disk and fast-moving broad line 
clouds within the inner 0.1 pc, surrounded by an optically-thick torus of dust and gas with narrow line 
clouds present at larger radii (see e.g.,\cite{1993ARA&A..31..473A,1995PASP..107..803U} for 
reviews).  The torus absorbs the x-ray, UV, and optical emission from the central engine and re-emits 
it in the infrared.  This makes mid-infrared (MIR) observations particularly useful for studying the torus 
and its surroundings through both imaging and spectroscopy.  
	
	While significant strides have been made in the understanding of the MIR emission of the "radio 
quiet" (RQ) population of AGN, much less work has been done with "radio loud" (RL) AGN.  About 
10-20\% of AGN have strong radio emission.  These RLAGN have powerful radio jets, as opposed to 
the larger RQ population, which have very small, low-power jets or none at all.  Jets emit via 
non-thermal processes, in particular via synchrotron radiation and (particularly at UV and higher 
frequencies) a variety of inverse-Compton mechanisms. This non-thermal emission dominates the 
properties of the source in the radio and mm, where the luminosity of RL AGN can be orders of 
magnitude higher than that of RQ AGN, but at near-infrared to X-ray wavelengths, RL and RQ AGN 
have remarkably similar broadband spectral properties (see e.g., \cite{Perlman2012}).  This indicates that, in particular, the optical and near-infrared emission of RL and RQ AGN are similar in nature, 
emerging from similar processes and regions of the AGN.  In the mid-to-far infrared, however, the
thermal emission from the torus must compete with emission from the high energy tail of the 
synchrotron component. 
 Simple power-law extrapolations of the radio and near-infrared spectra of RL 
AGN (e.g., \cite{impneug}) indicate that the crossing point for the thermal and 
non-thermal emission components should be somewhere in the infrared.  Thus in some objects
it is possible that the majority of the MIR emission may not come from warm dust in the 
AGN central regions (e.g., M87,  \cite{2001ApJ...561L..51P}), thus leaving open the question of the
MIR continuum's nature.

With this context in mind, then, mid-infrared observations of RL AGN at high angular resolution are of 
considerable importance.  Yet, to date, by far the majority of high angular resolution, MIR observations
of AGN are of RQ objects, with only a small number of RL objects having been observed.  
Recent MIR observations have started to fill in our knowledge of RL AGN.  MIR imaging of M87 shows 
a unresolved nucleus with a large radio jet with synchrotron emission from both the nucleus and 
extended jet \cite{2001ApJ...561L..51P, Whysong}, while the MIR imaging of Cen A shows an unresolved synchrotron 
nucleus surround by MIR emission from both dust and star-formation surrounding the nucleus 
\cite{2008ApJ...681..141R}.  The MIR imaging of Cen A also was able to place an upper limit of 4pc 
on the torus, which is consistent with findings from VLT interferometry \cite{2007A&A...471..453M}.  
MIR imaging of Cyg A revealed a complex picture, with a bright central source and a biconical, dusty 
emission region \cite{2002ApJ...566..675R}.  These structures resembled those seen with HST optical 
and near-IR 
observations \cite{1998MNRAS.301..131J,1999ApJ...512L..91T}.  These observations show the 
potential of high resolution MIR observations of RLAGN and highlights the need for these type of 
observations for a larger number of RLAGN.  Also, these three objects also have high resolution MIR 
spectroscopy observations \cite{2007ApJ...663..808P, 2007A&A...471..453M, 2000ApJ...535..626I, Merlo}.  

Other imaging observations are now being made, filling in one side of the story.  For example, 
a larger high resolution imaging survey of RL AGN was presented in 
\cite{2010A&A...511A..64V}.  Another 6, lower-luminosity objects were presented in \cite{Mason2012} 
as part of their work on mid-IR imaging of LINERs.    In this 
paper, we will present  multiband MIR imaging for six nearby (i.e z$<$0.1), bright (i.e. N band flux $>
$50 mJy in ISO and/or {\it Spitzer} observations) RL AGN.  
Along with 
the information gained from these observations, they will set the stage for future high resolution 
spectroscopy, allowing a complete picture of the objects to be obtained.      

\section{Results}
\label{sec:obs}

	MIR imaging observations were taken for six nearby RL-AGN using the 8.1m Gemini North and 
South telescopes and their respective mid-IR imaging cameras Michelle and T-ReCS.  
The objects were selected to have 12 $\mu$m flux $> 50$ mJy in ISO and/or {\it Spitzer} observations.
Table 1 lists critical information for these objects.  

For our observations, three different filters were used: the Si-2 filter centered at 8.7 $\mu$m 
with a bandwidth of 0.9 $\mu$m, the Si-6 filter centered at 12.3 $\mu$m with a bandwidth of 1.2 $\mu
$m, the Qa filter centered at 18.3 $\mu$m with a bandwidth of 1.5 $\mu$m. This allows us to observe both continuum and line emission, with a spatial resolution of $\sim 0.3''$, about 
10$\times$ greater than that of {\it Spitzer} ($\sim 3''$).  Typical observing times
were 10 minutes per band on-source, with the exception of Pictor A, where much deeper observations ($\sim 1.5$ hours on source) were obtained at 8.7 and 12.3 $\mu$m. In \cite{Merlo}
we also give the details of our observations and data reduction procedures.  We note that the upper limit given for M84 is at the  2 $\sigma$ level and agrees with the observations of \cite{Mason2012}.

\begin{table}[h]
\begin{center}
\begin{tabular}{llllc}
\hline\hline
Object Name & Redshift & F(8.7) & F(12.3) & F(18.3)\\
& & & mJy \\
\hline
3C84 &0.0176 & $372 \pm 4$ & $621 \pm 6$ & ... \\
3C120 & 0.0330 & $110 \pm 1$ & $226 \pm 2$ & ... \\
M84 &  0.00354 & $<1.1$ & $<8.8$ & ... \\
Cen B & 0.0129 & $35.1 \pm 0.6$ & $64 \pm 1$ & ... \\
Fornax A & 0.00587 & $8.1 \pm 0.5$ & $5.5 \pm 0.2$ & ... \\
Pic A & 0.0351 & $31 \pm 6$ & $54 \pm 1$ & $55 \pm 5$ \\
\hline
\end{tabular}
\end{center}
\caption{Our sample of RL-AGN.}
\label{tbl:obs}
\end{table}

Figure 1 shows images for three objects:  3C 84, 3C 120 and 
Pictor A.  The contour levels shown are based on the rms noise found in the images $
\sigma$, showing 1,2,3,4,5, 10, 15, and 20 $\sigma$ above background.  All of these images show a 
strong unresolved source, a common feature of all of our images.  In a few cases we find 
evidence of extended emission.  For example,  the 12.3 $\mu$m image of  3C 84 shows 
circularly symmetric extended flux at radii $>0.4''$.  We show the radial profile of that object in 
Figure 2, compared to the profile for the PSF standard observed immediately before.   This emission
could be due to circumnuclear star formation, 
given the H$\alpha$ filaments known in 3C 84's host galaxy and the presence of PAH lines 
in the {\it Spitzer} spectrum \cite{Leipski2009}.  Deeper observations are required to untangle this 
subject.   

\begin{figure}[t]
\includegraphics[scale=0.75]{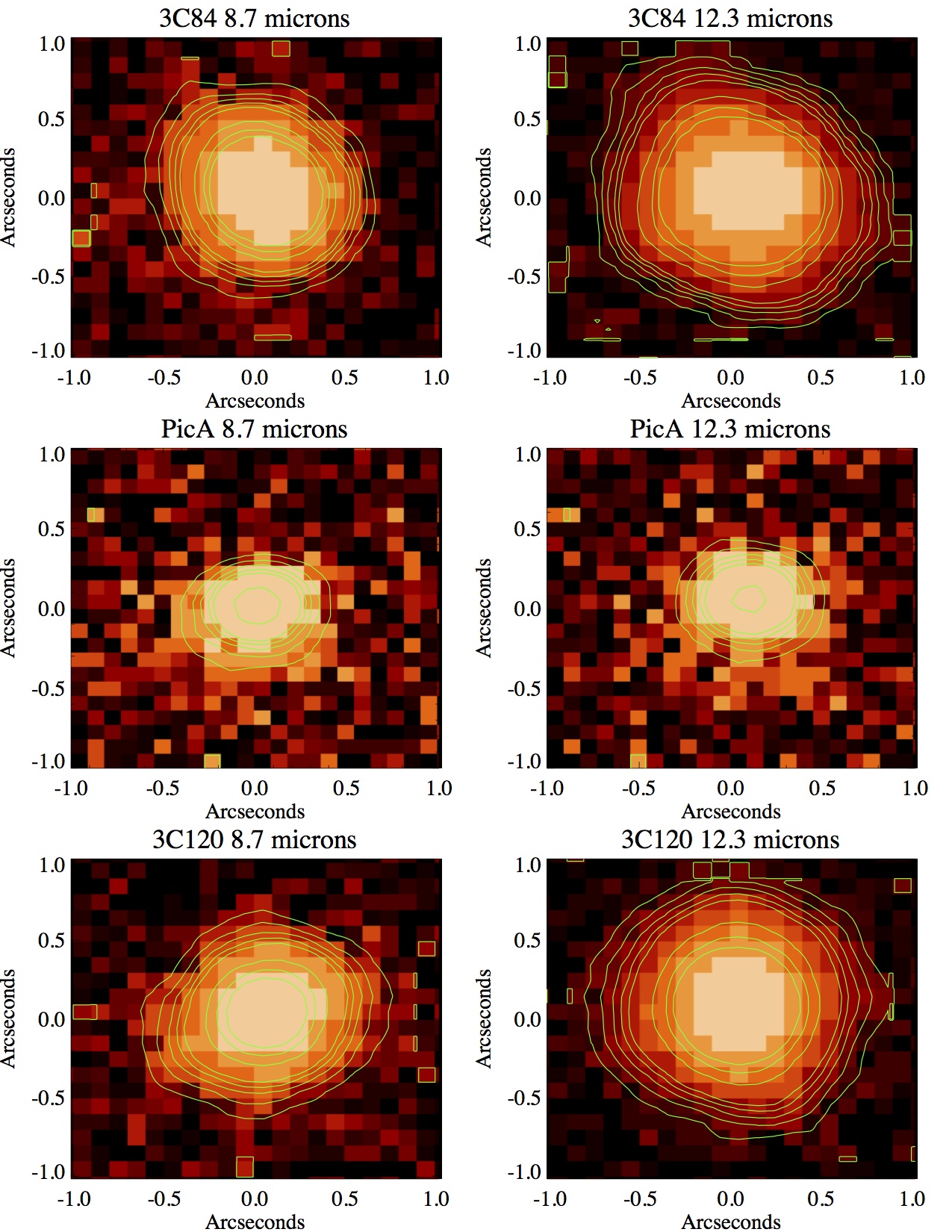}
\caption{Mid-IR images of 3C 84, 3C 120 and Pic A at 8.7 $\mu$ m and 12.3 $\mu$ m.  Contours 
are plotted at  1, 2, 3, 4, 5, 10, 15 and 20 $\sigma$ above the background.}
\end{figure}

\section {Correlation between X-ray and MIR Luminosity and Discussion}

\label{sec:midirvsxray}

The MIR (12 $\mu$m) and hard X-ray (2-10 keV) luminosities of AGN are well correlated, as first 
reported by \cite{Krabbe2001} and \cite{Lutz2004} based on ISO data.  
This correlation was elaborated on using ground-based data by \cite{Horst2008}, and refined
by \cite{Levenson2009} and \cite{Gandhi2009}
who used data for objects with $<100 pc$ resolution as well as absorption-corrected X-ray luminosity.
This correlation is beginning to be explored for radio loud objects as well.    
\cite{2011A&A...536A..36A} added the radio-loud AGN observed to date to that correlation.  Here we 
add the radio-loud AGN of this work and also \cite{Mason2012} and \cite{2010A&A...511A..64V} 
to that plot, using the  
mid-IR luminosity within a 100 pc region where available, as in \cite{Levenson2009}.  
This plot is shown in Figure 3.

\begin{figure}[t]
\includegraphics[scale=0.75]{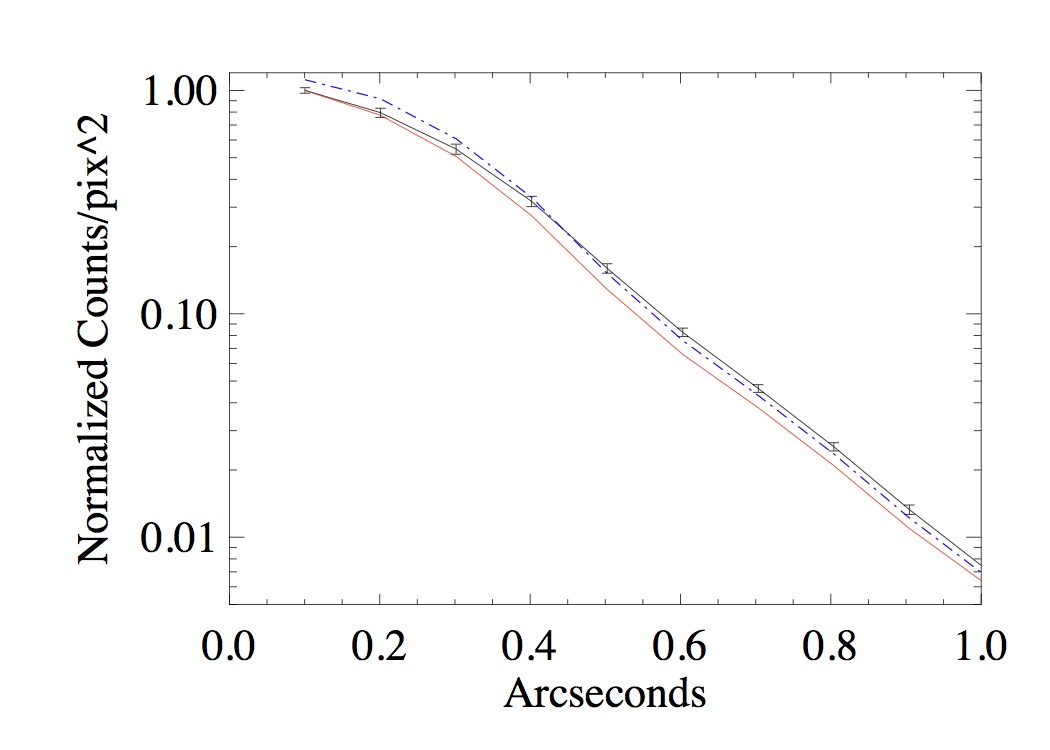}
\caption{The radial profile for 3C 84 at 12.3 $\mu$ m (points, with 1$\sigma$ error ranges shown, and black curve).
Plotted for comparison are the PSF as 
derived from a flux calibration star observed the same night (red curve), as well as a curve at $3 
\sigma$ above the PSF (dash-dot curve). }

\bigskip
\bigskip

\includegraphics[scale=0.75]{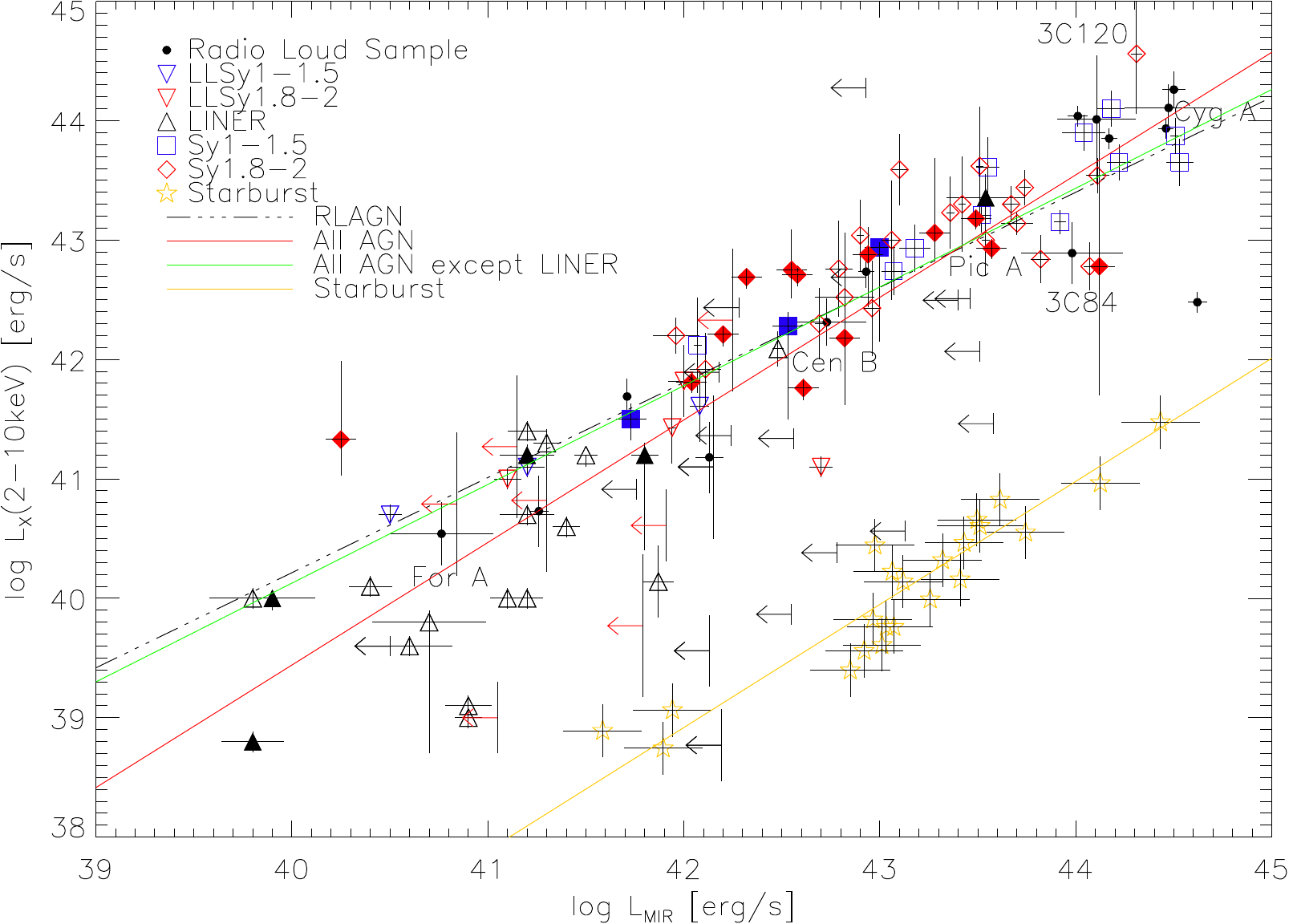}
\caption{The mid-IR vs. X-ray luminosity correlation for radio-loud as well as radio-quiet AGN.  See Section 3 for discussion.}
\label{fig:midirvsxray}
\end{figure}

As can be seen, the RL AGN do not differ significantly from RQ AGN.  The RL AGN show a highly 
significant correlation, with a Spearman rank order correlation coefficient, $\rho$=0.797, 
corresponding 
to a null hypothesis probability of p=6.28$\times$10$^{-4}$.   There is no significant difference 
between the slopes of the two classes.   While this is less secure than for all the AGN plotted ($
\rho=0.926, p=1.69 \times 10^{-38}$), this is purely  a result of the smaller number of RL AGN 
observed.
The fact that this correlation seems to hold for both RL AGN and RQ QGN indicates that the X-ray 
luminosity can be linked to that in the mid-IR in a similar way for all AGN, regardless of radio 
loudness.  Our results also indicate that for most RL AGN, the mid-IR band is dominated by
physical processes similar to those seen in RQ AGN, i.e., thermal emission from the torus 
and/or star formation, 
with a more minor contribution from non-thermal synchrotron emission -- although exceptions such
as M87 certainly exist.  The only difference between classes appears to be for the LINERs, which 
do appear to skew the correlation somewhat.  We are investigating this last observation further.

\cite{Horst2008} and \cite{Levenson2009} interpreted the correlation as evidence for a 
compact, clumpy torus, as opposed to a much larger one that is smooth in morphology 
\cite{KroBeg88}.  A larger but smooth torus would also have correlated MIR and X-ray luminosities, 
but with a larger mean value of $L_{MIR}/L_{X}$ for Sy 1s than Sy 2s, which is not seen in 
RQ AGN (e.g., \cite{Levenson2009, Gandhi2009, Horst2008}).  While the RL AGN are also 
consistent with no differences between the Sy 1 and Sy 2 correlation, this statement cannot be
made securely as there are too many upper limits on the MIR luminosity (particularly in the work of 
\cite{2010A&A...511A..64V}).  The data thus lead us to infer that the tori of RL AGN are of roughly
similar size and morphology as those of RQ AGN -- namely, being sub-parsec in size.  This is a 
significantly tighter constraint than we can gain using our imaging data alone, where due to the 
greater distances of the objects we can only constrain the size of the tori to $\sim 10-50$ pc.  We 
will 
explore this issue in a later paper.

\end{document}